\documentclass[11pt,twoside]{article}
\usepackage{./asp2014}

\aspSuppressVolSlug
\resetcounters

\begin{document}

\title{Discovery of a Centrifugal Magnetosphere Around the He-Strong Magnetic B1 Star ALS 3694}
\author{Matt Shultz,$^{1,2,3}$ Gregg Wade,$^3$ Thomas Rivinius,$^1$ James Sikora,$^{2,3}$ and the MiMeS Collaboration
\affil{$^1$European Southern Observatory, Santiago, Chile; \email{mshultz@astro.queensu.ca}}
\affil{$^2$Queen's University, Kingston, Ontario, Canada}
\affil{$^3$Royal Military College, Kingston, Ontario, Canada}}

\paperauthor{Matt~Shultz}{mshultz@astro.queensu.ca}{}{Queen's University}{Dept. of Physics, Engineering Physics, and Astronomy}{Kingston}{ON}{K7L 3N6}{Canada}
\paperauthor{Gregg~Wade}{Gregg.Wade@rmc.ca}{}{Royal Military College}{Dept. of Physics}{Kingston}{Ontario}{K7K 7B4}{Canada}
\paperauthor{Thomas~Rivinius}{triviniu@eso.org}{}{European Southern Observatory}{Astronomy}{Vitacura}{Santiago}{19001}{Chile}
\paperauthor{James~Sikora}{james.sikora@queensu.ca}{}{Queen's University}{Dept. of Physics, Engineering Physics, and Astronomy}{Kingston}{ON}{K7L 3N6}{Canada}

\begin{abstract}
We report the results of 6 nights of Canada-France-Hawaii Telescope spectropolarimetric ESPaDOnS observations of the He-strong, magnetic B1 type star ALS 3694. The longitudinal magnetic field is approximately 2 kG in all 6 observations, showing essentially no variation between nights. The H$\alpha$ line displays variable emission on all nights, peaking at high velocities ($\sim 3 v\sin{i}$). Given the presence of a strong ($B_{\rm d}>$6 kG) magnetic field, and the similarity of the emission profile to that of other magnetic B-type stars, we interpret the emission as a consequence of a centrifugal magnetosphere. 
\end{abstract}

\section{Introduction}

ALS 3694 is a poorly studied star in the young open cluster NGC 6193 \citep{landstreet2007}. It is relatively dim ($V = 10.4$), near the magnitude limit for high-resolution spectropolarimeters. Two FORS1 circularly polarized (Stokes $V$) observations have been reported by \cite{bagnulo2006}. A magnetic field was detected in both observations, with the mean longitudinal (line-of-sight) magnetic field $B_{\rm Z} = -1.9 \pm 0.2$ kG.  

\section{Observations}

We have acquired 6 nights of ESPaDOnS observations of ALS 3694 with the Canada-France-Hawaii Telescope (CFHT). ESPaDOnS is a high-resolution ($\lambda/\Delta\lambda\sim 65,000$) spectropolarimeter, with a spectral window from 369 nm to 1050 nm. Each observation consists of four polarized sub-exposures, which are combined via the double-ratio method to yield a Stokes $I$ (unpolarized intensity) and Stokes $V$ spectrum \citep{d1997}. The 4 sub-exposures can also be combined in such a way that the intrinsic source polarization cancels out, yielding a diagnostic null $N$ spectrum \citep{d1997} which can be used to check for spurious Stokes $V$ signatures arising at the instrumental level, or from radial velocity variation of the source due to binary motion or pulsation (e.g. \citealt{neiner2012}). 

The large visual magnitude of this star necessitated long integration times, 20 min per spectropolarimetric sequence. Between 2 and 4 successive sequences were collected each night (16 individual polarized spectra in total). These were binned nightly. The mean peak SNR of the binned spectra is 145. 

\articlefiguretwo{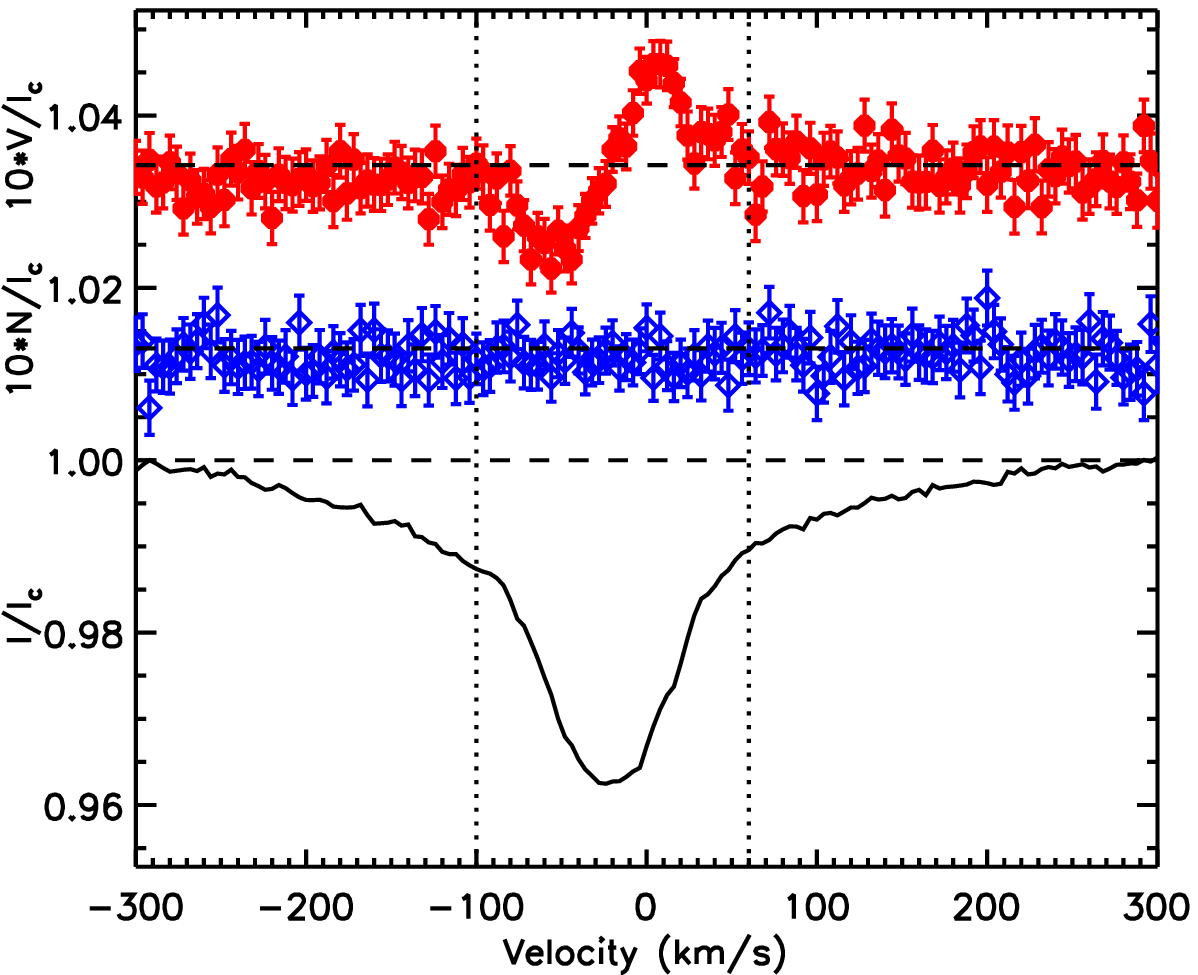}{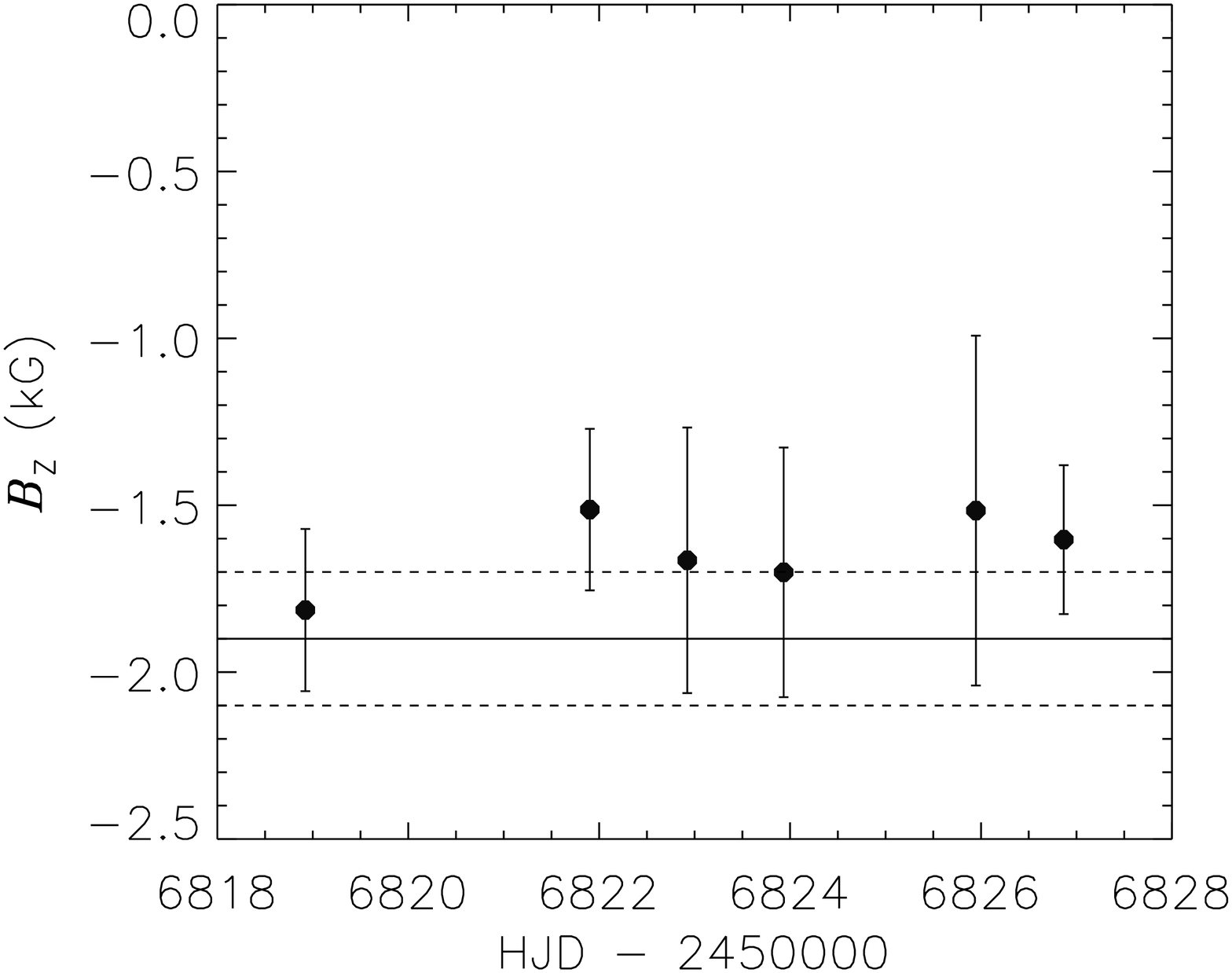}{lsd_bz}{{\em Left}: An example LSD profile. Top (filled circles), Stokes $V$; middle (open diamonds) diagnostic null $N$; bottom (solid line) Stokes $I$. Vertical dotted lines indicate the integration range used for measurement of the longitudinal magnetic field. Horizontal dashed lines indicate continuum levels. {\em Right}: $B_{\rm Z}$ measurements as a function of HJD. There is no variation from between nights outside the error bars. The solid and dotted horizontal lines indicate the mean $B_{\rm Z}$ and the 1$\sigma$ uncertainties from the two FORS1 measurements.}

\section{Magnetic Field}

In order to increase the SNR, we performed Least Squares Deconvolution (LSD; \citealt{d1997}), using the `improved' iLSD package provided by \cite{koch2010}. This consists of deconvolving a mean line profile (the LSD profile) from the spectrum using a `line mask': a series of delta functions at the laboratory wavelengths of the spectral lines, each weighted by the depth of the line, and the Land\'e factor (a dimensionless measure of the magnetic sensitivity of the line). 

The line mask was downloaded from the Vienna Atomic Line Database, VALD2 \citep{piskunov1995, ryabchikova1997, kupka1999, kupka2000}, created for a star with $T_{\rm eff} = 22$ kK, $\log{g} = 4.0$, and solar metallicity. The raw line mask was cleaned of strong, pressure-broadened H lines, since these lines have a substantially different shape from the majority of photospheric absorption lines. Also removed were any He and metallic lines blended with H lines, interstellar lines, or  telluric lines. While the usual cleaning procedure also removes the stronger He lines (since these are subject to pressure-broadening), in this case all such lines were included, as their contribution to increasing the SNR of the LSD profile outweighs their distortion of the Stokes $I$ profile. The depths of the remaining lines were then empirically adjusted (`tweaked') so as to match the observed depths of the absorption lines (see \cite{shultz2012} for an explanation of this process).

An example LSD profile is shown in Fig. \ref{lsd_bz}. Stokes $I$ shows extended wings due to the inclusion of strong He lines. The Zeeman signature of a magnetic field is clearly seen in Stokes $V$, while the $N$ profile is consistent with noise. 

The line-of-sight (longitudinal) magnetic field was measured from the LSD profiles using zeroth moment of Stokes $V$ normalized by the equivalent width of Stokes $I$, as described by \cite{mat1989}. $B_{\rm Z}$ is shown in Fig. \ref{lsd_bz} as a function of HJD. There is no variation outside error bars from night to night. Furthermore, all measurements are consistent with the earlier FORS1 data \citep{bagnulo2006}.

\section{H$\alpha$ Emission}

\articlefiguretwo{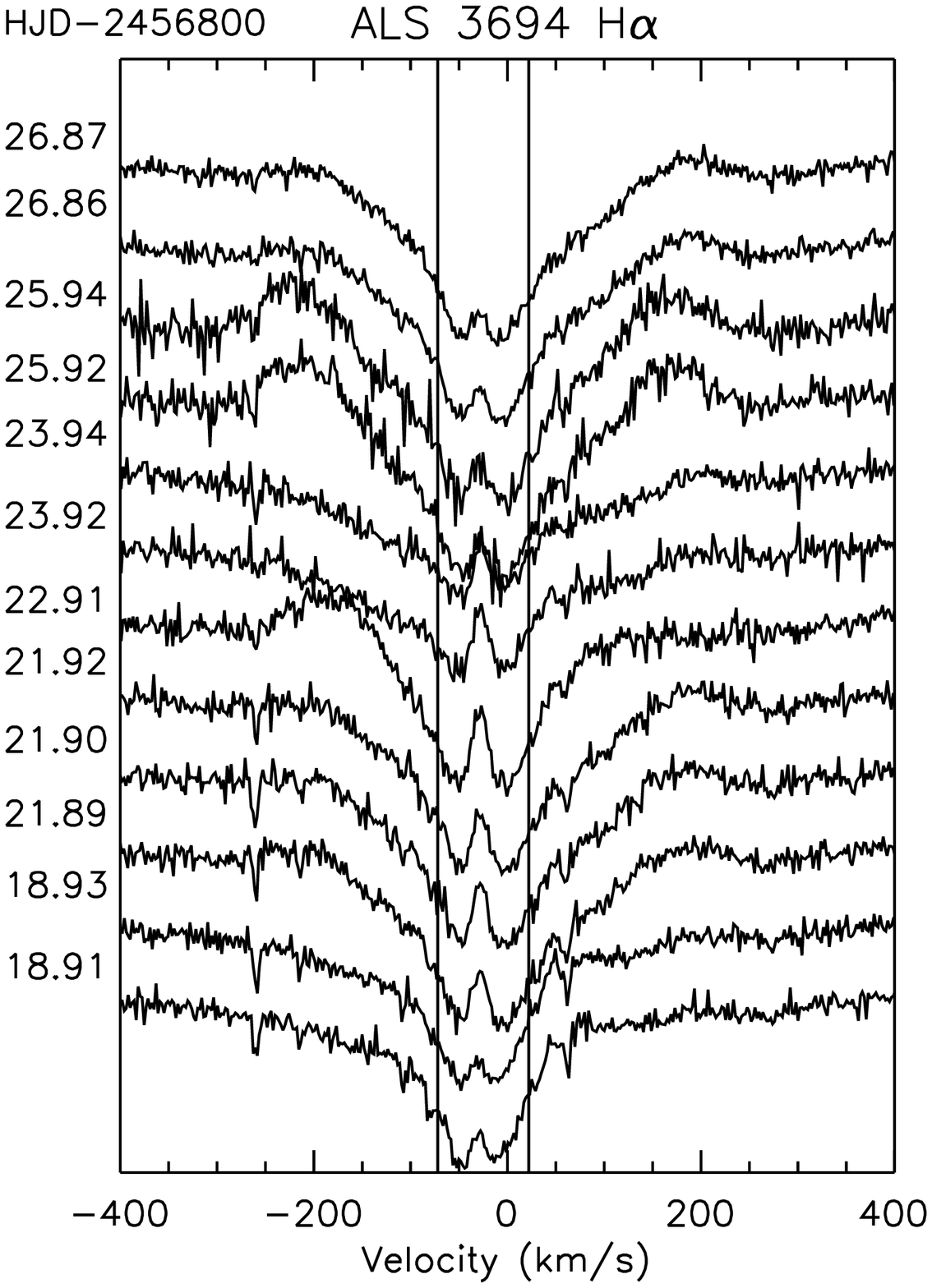}{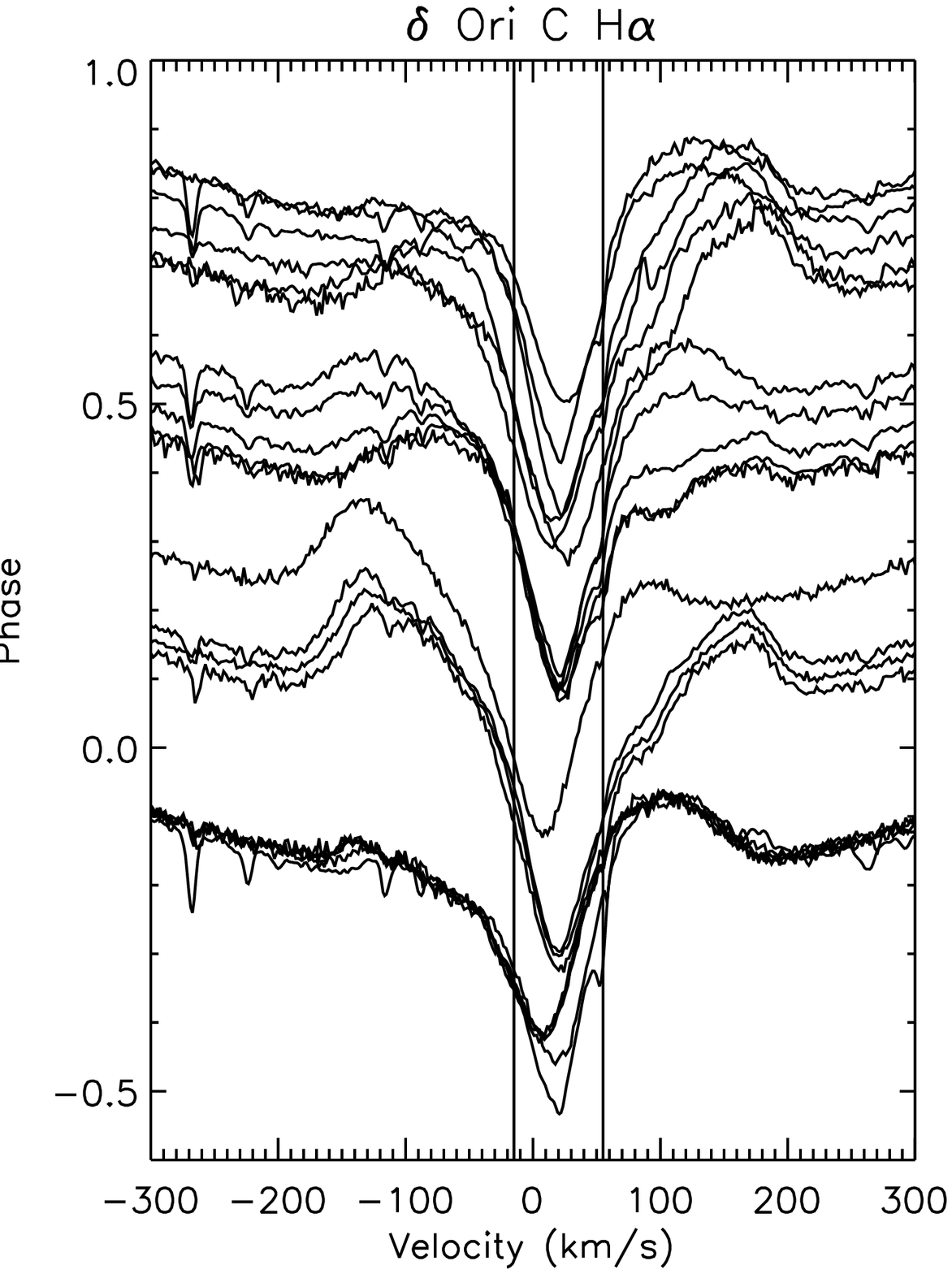}{halpha}{H$\alpha$ profiles. Solid vertical lines indicate $\pm v\sin{i}$, centred on the stellar systemic velocity. {\em Left}: ALS 3694 arranged in order of HJD. There is substantial night-to-night variability. {\em Right}: $\delta$ Ori C as a function of rotational phase.}

Fig. \ref{halpha} shows H$\alpha$ profiles for ALS 3694, and for the magnetic B3 Vp star $\delta$ Ori C (note that $\delta$ Ori C is an SB2 \citep{leone2010}, and has not been corrected for radial velocity variations). In both cases, emission is present in essentially all observations, either double- or single-peaked, with peaks in either the red- or blue-shifted wings of the lines. Vertical lines indicate $\pm v\sin{i}$: in both cases, emission peaks at $\sim 2-3 v\sin{i}$. Note that we have re-measured $v\sin{i}$ for ALS 3694 using He, N, O, Ne, Si, and S lines, finding $v\sin{i} = 47 \pm 4 {\rm km~s}^{-1}$, substantially lower than the  82 ${\rm ks~s}^{-1}$ reported by \cite{landstreet2007}. 

\section{Discussion}

While the rotational period $P_{\rm rot}$ is unknown, $P_{\rm rot} < 1.5$ d for every known example of a magnetic B-type star with circumstellar emission (Shultz et al., these proceedings). Proceeding on the assumption that ALS 3694 is no exception, we estimae that the inclination $i$ of the rotational axis from the line of sight must be in the range $5^\circ < i < 21^\circ$, where the lower bound comes from the requirement that the equatorial rotational velocity must be less than the breakup velocity. 

Adopting the method of \cite{preston1967} for determining the angle $\beta$ between the magnetic and rotational axes and the strength of the magnetic dipole $B_{\rm d}$, we find that $15^\circ < \beta < 63^\circ$, and $B_{\rm d} > 7$ kG. A small $i$ is also supported by the lack of variation in $B_{\rm Z}$ (although of course, it is also possible that $i$ is large and $\beta$ is small). This arrangement, with a small $i$, moderate $\beta$, and substantial $B_{\rm d}$, is reminiscent of that of $\delta$ Ori C ($i = 12 \pm 3^\circ$, $\beta < 52^\circ$, $B_{\rm d} = 9.7 \pm 2.4$ kG; \citealt{leone2010}). 

Using the stellar parameters ($T_{\rm eff} = 20 \pm 3$ kK, $\log{(L_*/L_\odot)} = 3.7 \pm 0.3$) given by \cite{landstreet2007}, and the mass-loss recipe of \cite{vink2001} (yielding $\log{\dot{M}} = 9 \pm 1$ and $v_\infty = 1750 \pm 750 {\rm km~s}^{-1}$), we estimate the Alfv\'en radius $R_{\rm A}$ (that is, the maximum extent of closed magnetic loops) to be $R_{\rm A} > 20 R_*$ \citep{ud2002}. The Kepler radius $R_{\rm K}$ (the radius at which centrifugal and gravitational forces balance; see \citealt{town2005}, \citealt{ud2008}, also Shultz et al., these proceedings) can be estimated from $v\sin{i}$, $R_*$, and the likelihood that $i$ is large: we then have $R_{\rm K} < 2.9 R_*$. 

Given that $R_{\rm A} >> R_{\rm K}$, and the strong resemblence of ALS 3694's H$\alpha$ emission to that of other magnetic B-type stars with detected magnetospheres, it seems very likely that ALS 3694 is the most recent example of the growing sub-class of magnetic B-type stars with centrifugal magnetospheres.

\bibliography{bib_dat}  

\end{document}